\documentclass[runningheads]{llncs}
\usepackage[T1]{fontenc}

\usepackage{amsmath,amssymb}

\usepackage{xcolor}
\usepackage{graphicx}

\usepackage{listings}

\lstdefinestyle{mystyle}{
    basicstyle=\ttfamily\footnotesize,
    breakatwhitespace=false,         
    captionpos=b,                    
    keepspaces=true,                 
    numbers=left,                    
    numbersep=5pt,                  
    showspaces=false,                
    showstringspaces=false,
    showtabs=false,                  
    tabsize=2
}

\usepackage{hyperref}
\usepackage{cleveref}
\usepackage{multicol}

\usepackage[colorinlistoftodos]{todonotes}

\title{Crux, a Precise Verifier for Rust and \\ Other Languages}
\author{Authors anonymised}
\authorrunning{Anon et al.}
\institute{Institutions anonymised}

\author{
    Stuart Pernsteiner\inst{1}
    \and Iavor S. Diatchki\inst{1}
    \and Robert Dockins\inst{2}
    \and Mike Dodds\inst{1}
    \and Joe Hendrix\inst{2}
    \and Tristan Ravich\inst{2}
    \and Patrick Redmond\inst{2}
    \and Ryan Scott\inst{1}
    \and Aaron Tomb\inst{2}
}
\authorrunning{S. Pernsteiner et al.}

\institute{Galois, Inc. \and Work performed while at Galois, Inc.}

\begin{document}
\maketitle

\begin{abstract}
We present \emph{Crux} a cross-language verification tool for Rust and C/LLVM. 
Crux targets bounded, intricate pieces of code that are 
difficult for humans to get right: for example, cryptographic modules
and serializer / deserializer pairs. Crux builds on the same framework as
the mature SAW-Cryptol toolchain, but Crux provides an 
interface where proofs are phrased as symbolic unit tests. Crux is designed
for use in production environments, and has already seen use in industry.

In this paper, we focus on Crux-MIR, our verification tool for Rust.
Crux-MIR provides a bit-precise model of safe and unsafe Rust which can
be used to check both inline properties about Rust code, and extensional
equality to executable specifications written in Cryptol or in the
hacspec dialect of Rust. Notably, Crux-MIR supports compositional
reasoning, which is necessary to scale to even moderately complex
proofs. We demonstrate Crux-MIR by verifying the Ring library implementations of
SHA1 and SHA2 against pre-existing functional specifications.

Crux is available at \url{https://crux.galois.com}. 

\end{abstract}

\definecolor{LightGrey}{gray}{0.95}

\section{Introduction}
\label{sec:introduction}

Many types of systems can be verified, but in this paper, we are
interested in one kind in particular: intricate, highly optimized, and
encapsulated behind interfaces with precise algorithmic specifications.
These properties describes several important and security-critical
categories of system, in particular \emph{cryptographic modules} such as
AES, SHA, ECDSA, and others; and \emph{serializers/deserializers} such
as those found at network interfaces. In both cases, code is performance
and security-critical, with errors leading to costly consequences. Our 
aim is to formally verify this code in production. 


Recent work has demonstrated the viability of \emph{compositional symbolic
simulation} as an approach to formally verifying bounded, intricate code of this
kind. This approach is embodied in the mature SAW-Cryptol toolchain, which
comprises SAW~\cite{saw}, the verification tool, and Cryptol~\cite{cryptol}, the
specification language. This approach to verification has three interrelated
parts:

\begin{itemize}

\item
  \emph{Precise symbolic execution} performed by the Software Analysis Workbench
  (SAW). This transforms highly optimized production code into
  mathe\-matically-equivalent symbolic terms, represented in an SMT-like format. 

\item
  \emph{Verification of extensional equality} between the symbolic term
  representing the program and an executable specification. By extensional
  equality, we mean (1)
  that code and specification cannot generate undefined behaviour / panic, 
  and (2) that they produce 
  equivalent output when supplied with equivalent
  inputs\footnote{The proof must define the mapping between specification and implementation
  data, of course.}. 

  \item
  \emph{Compositional reasoning}, meaning that complex sub-functions can
  be overridden by simpler specification functions. Compositional reasoning 
  is necessary to scale to even moderately complex proofs in our domain. 

\end{itemize}

This approach targets \emph{mostly}-bounded
code. A high degree of automation is possible when loops have static
upper-bounds and data-structures are statically allocated and have fixed sizes.
If these properties do not hold in certain limited ways, loop invariants can be
used to enable verification, or arbitrary input bounds can be imposed, 
which reduces the power of the resulting theorems. However, the 
overwhelming majority of code in our target domain of cryptographic
primitives is inherently bounded, which enables a high degree of automation. 

Compositional symbolic simulation has been quite successful in formally
verifying cryptographic code in industry. Most recently, a team at Galois
and Amazon Web Services verified the s2n and AWS-LibCrypto libraries using
SAW-Cryptol~\cite{continuous-verification-s2n,Boston2021,DBLP:journals/ieeesp/Dodds22}.
These are production-grade cryptographic libraries written in C and x86 assembly, which
were verified without code modifications. 

\paragraph{Why Crux?}
\label{sec:crux_intro}

Crux\footnote{\url{https://crux.galois.com}} is a new tool which presents an alternative interface to compositional
symbolic simulation. In doing this, it aims to improve over SAW-Cryptol as a
production verification tool. In SAW-Cryptol 
proofs must be written in two custom DSLs, SAW-script and Cryptol. This presents
a significant learning curve, even for seasoned experts.
Crux addresses this in two ways. First, proofs are written in the
form of unit tests rather than in SAW-Script (proofs written in this
form are called \emph{symbolic tests}). Second, specifications can be written in the native
language---in this case Rust---rather than the Cryptol DSL. For example,
consider the following two pieces of Rust code:

\noindent 
\begin{minipage}[t]{0.45\linewidth}
\begin{lstlisting}
#[test]
fn f_output_natural(){
    let x = 42;

    let y = f(x);
    assert!(0 <= y);
}
\end{lstlisting}
\end{minipage}
\begin{minipage}[t]{0.55\linewidth}
\begin{lstlisting}
#[crux_test]
fn f_output_natural2(){
    let x = <i32>::symbolic("x");
    crucible_assume!(-100 <= x);
    let y = f(x);
    crucible_assert!(0 <= y);
}
\end{lstlisting}
\end{minipage}

On the left, a Rust unit test asserts that the function
\lstinline{f} outputs a natural number for the concrete input
\lstinline{42}. On the right, we have converted the unit test to
a Crux-MIR symbolic test. This formally verifies that the result is a
natural number for all inputs in the range
\lstinline{-100_i32..i32::MAX}. We describe the syntax of Crux 
proofs later in the paper.

This style of proof is familiar from tools such as
CBMC~\cite{cbmc}, and is a standard interface for the popular SV-COMP
 competition~\cite{SVCOMP22}. What is novel about Crux is that
it combines this proof style with the advantages of SAW-Cryptol, namely
compositional reasoning, support for Cryptol specifications,
and bit-precise verification of intricate
bounded code.

\paragraph{Contributions.}

Our first contribution is Crux itself. Crux is designed as a cross-language tool
(see \emph{Other Languages}, below) but in this paper, we focus on the Rust
front-end, Crux-MIR. This is a bit-precise verification tool which can
reason about safe and unsafe Rust code, and which we have already used in production. 
 
Crux-MIR's proof interface is novel, in that it combines a symbolic testing
interface as seen in CBMC and other tools, with compositional symbolic
simulation as seen in SAW-Cryptol, and cross-language proofs of extensional
equality. Crux is the first automated verification tool to support the hacspec
specification language, as well as the domain specific language Cryptol. Crux is not 
intended for every task. Rather, it is a particular point in the
design space which addresses our needs when verifying cryptographic
libraries and similar. 

Crux-MIR has already been used as part of an industrial deployment at Amazon 
Web Services---see \cref{sec:shardstore} below. 
In this paper, we demonstrate Crux-MIR by verifying the Ring library
implementations of SHA1 and SHA2~\cite{ringlibrary} against executable specifications
written in hacspec and Cryptol. To the best of our knowledge these
proofs are the first verification of library cryptography primitives implemented 
solely in Rust. 
The proofs are not trivial, but nor are they as complex as recent the industry proofs of s2n
and AWS-LibCrypto. Rather, they are intended to demonstrate Crux-MIR's capabilities
when verifying off-the-shelf code not originally designed for verification.


Crux is built on the same infrastructure used by SAW-Cryptol, despite
its distinct proof style. The core of Crux is a symbolic
execution library called \emph{Crucible} which is described for the first
time in this paper\footnote{Crucible is also the back-end library for SAW-Cryptol and several 
other tools.}. Crucible provides an abstracted framework for
performing symbolic execution on imperative programming languages. Our
final contribution for this paper lies in demonstrating how Crucible
and other related libraries make it easy to build new language models (in this
case Rust) and new proof interfaces (in this case, Crux itself). They provide a
toolbox of useful components that can be used to construct new 
verification tools. 

\paragraph{Other languages.}

Crux is designed as a cross-language verification tool. 
As well as Crux-MIR, it supports two other interfaces.
These share the same basic architecture as Crux-MIR. 
The program semantics is encoded into the cross-language Crucible IR, which 
is then translated into logical assertions by symbolic execution. 
\begin{itemize}

    \item Crux-LLVM: this interface supports C and C++ code compiled to LLVM, at
    parity with the C/LLVM support in SAW-Cryptol. Using this interface, a team
    was recently able to submit Crux to the SV-Comp competition in several
    categories~\cite{crux-sv-comp}.  

    \item Crux-x86 / Crux-ARM: these interfaces are in development. At present, 
    we have modeled the semantics of x86-64 and ARM-32, and can verify 
    test examples. However, we do not have a proof interface that can 
    set up states via assertions---this must be done by injecting custom code. 

\end{itemize}

\section{Crux by Example: Vector Clocks and ChaCha20}
\label{sec:example}

We introduce Crux with two examples. First, vector clocks---logical clocks that
model the partial causal order of events~\cite{10.1145/10590.10593}. We use this
simple example to demonstrate the structure of Crux proofs and the way that
Crux compositional reasoning is applied. We then examine another simple but more
realistic example, the ChaCha20 cryptographic primitive (see
\cref{sec:example2}).

\subsection{Vector Clocks: Verifying Commutativity}

\begin{figure}
    \begin{lstlisting}
const N: usize = 8;
type Clock = u32;
type VectorClock = [Clock; N];

fn merge_clocks(a: Clock, b: Clock) -> Clock {
    a.max(b)
}
fn merge_vc(a: VectorClock, b: VectorClock) -> VectorClock {
    let mut out = [0; N];
    for i in 0..N {
        out[i] = merge_clocks(a[i], b[i]);
    }
    out
}
    \end{lstlisting}
    \caption{
        \label{fig:example-impl}
        Vector clock type and associated merge implementation in Rust.
    }
\end{figure}

First, Vector clocks. For illustrative purposes, we only
examine the vector clock merge function, and two correctness
properties: (1) commutativity of merge operations; and (2) input-output
correctness of the merge function with respect to a functional
specification.

A vector clock consists of a list of natural numbers that count the
number of events observed at each of \(N\) sites.
\Cref{fig:example-impl} shows a Rust implementation of a vector clock
type and its associate merge function. The merge function computes the
element-wise maximum of event counts. We
define the trivial \lstinline{merge_clocks} sub-function for use
in demonstrating compositional reasoning.

\begin{figure}
\begin{lstlisting}
#[cfg(crux)]
mod crux_commutativity {
    // ...
    #[crux_test]
    fn merge_vc_commutative() {
        let a = <VectorClock>::symbolic("a");
        let b = <VectorClock>::symbolic("b");
        let ab = merge_vc(a, b);
        let ba = merge_vc(b, a);
        crucible_assert!(ab == ba);
    }
}
\end{lstlisting}
\caption{
    \label{fig:example-comm}
    Crux symbolic test specifying that \lstinline{merge_vc} is commutative.
}
\end{figure}

Expressed mathematically, we want the following commutativity property
to hold of vector clocks. This property helps ensure that merging 
is insensitive to network delays; the same value is always
arrived at, no matter the ordering. \[
\forall a, b \in \mathtt{VectorClock}. \;
            \mathtt{merge\_vc}(a, b)
    \equiv \mathtt{merge\_vc}(b, a)
\] \Cref{fig:example-comm} expresses this commutativity property of
\lstinline{merge_vc} using Crux's proof format, that of a
\emph{symbolic test}. Test modules are marked with the
\lstinline{#[cfg(crux)]} attribute, and test functions are marked
with the \lstinline{#[crux_test]} attribute.

First, we declare two \emph{symbolic} vector clocks values by calling
the trait method \lstinline{Symbolic::symbolic(&'static str)}
\emph{(lines 6-7)}. Crux provides implementations of
\lstinline{Symbolic} for common Rust standard library types, and
so it works for our vector clock type alias. The string parameters
\lstinline{"a"} and \lstinline{"b"} are arbitrary names
that are used for debugging a failed test.

A symbolic test can constrain the possible values of a symbolic value
using the macro \lstinline{crucible_assume!($e:expr)}. This has
the effect of imposing a precondition on the values of the symbolic
variables. This macro can be seen in the
\lstinline{f_output_natural2} example in \Cref{sec:crux_intro},
where it is used to assert that the variable \lstinline{x} is a
natural number.

Next we call the system under test, \lstinline{merge_vc}, once
with each argument order to generate the results we expect are equal
despite the swapped order, according to the definition of commutativity
\emph{(lines 8-9)}. Finally, a call to
\lstinline{crucible_assert!($cond:expr)} asserts that the results
are equal \emph{(line 10)}. This has the effect of requiring a
post-condition on the values of symbolic variables.

As we observed in the introduction, there is a strong resemblance
between a symbolic test and a concrete unit test. In a concrete version
of this program, the variables \lstinline{"a"} and
\lstinline{"b"} would be instantiated with particular vector
clocks, and the \lstinline{crucible_assert!($cond:expr)} would be
replaced with a standard assertion. However, the structure of the test
would be the same. See Kellogg~\emph{et al} for a
discussion of how this proof style can be appealing to domain expert
engineers in production~\cite{continuous-compliance}.

Crux similarly intentionally mirrors the normal
Rust testing workflow. To execute the symbolic test, the user runs
\texttt{cargo\ crux-test}. This attempts to identify an assignment of
the symbolic values to concrete values which adheres to the constraints
expressed by \lstinline{crucible_assume!} but violates the
constraints expressed by \lstinline{crucible_assert!}. If none is
found, the symbolic test passes, meaning the theorem is proved. If the
test fails, Crux can provide a counter-example.

\subsection{Vector Clocks: Specifying Functional Properties}
\label{sec:example-spec}

The commutativity of \lstinline{merge_vc} has a relatively simple
specification which can be expressed by asserting the equality of return
values. This follows one common pattern during
test-driven-development---inline Boolean assertions on values. However,
more complex functional properties of code often cannot be specified in
this way. This specification does not establish that the vector clock
contains the correct values, but just that the operations commute. To
specify these internal properties, a proof tool must have some way of
specifying the desired functional behavior of the code.

Crux handles these cases by verifying that code is \emph{extensionally
equal} to a reference implementation written in either the current
language (i.e Rust / hacspec) or in some higher level specification language 
or in the high-level Cryptol specification DSL. We
will demonstrate this functionality with Cryptol, and discuss Rust
support later. 

\begin{figure}
    \begin{lstlisting}
mod cryptol_spec {
    // ...
    cryptol! {
        pub fn merge_clocks(a:Clock,b:Clock) -> Clock
            = r#"\(a: [32]) (b: [32]) ->
                    if a > b then a else b"#;
        pub fn merge_vc(a: VectorClock, b: VectorClock) -> VectorClock
            = r#"\(as: [8][32]) (bs: [8][32]) ->
                    [ if a > b then a else b
                        | a <- as
                        | b <- bs ]"#;
    }
}
    \end{lstlisting}
    \caption{
        \label{fig:example-spec}
        Cryptol specification for use with Crux-MIR verification.
    }
\end{figure}

\Cref{fig:example-spec} shows the specification for
\lstinline{merge_vc}. Cryptol is a Haskell-like language so the
merge function can be specified with a list comprehension \emph{(lines
8-11)}. This specification captures exactly the desired functional
behavior for a merge function, namely computing the element-wise
maximum of event counts. We could equally well specify this function in
a piece of executable Rust, albeit in this very simple case the
specification would be very similar to the implementation.

Cryptol is an executable language, so this specification can also be
used to test inputs and outputs, and it is designed to be easy to audit
for experts, especially for cryptographic modules. The example we show
here is the \emph{inline} approach to defining specifications. In this
case, the Cryptol bindings are in Rust's name-space, and so they cannot
reference each other. For more complex examples, we also support
importing Cryptol from separate files, in which case programs can
include richer structure.

\subsection{Vector Clocks: Verifying Correctness, Compositionally}
\label{sec:example-equiv}

\begin{figure}
\begin{lstlisting}
#[cfg(crux)]
mod crux_spec_equivalence {
    // ...
    #[crux_spec_for(merge_clocks)]
    fn merge_c_equiv() {
        let a = <Clock>::symbolic("a");
        let b = <Clock>::symbolic("b");
        let out = merge_clocks(a, b);
        crucible_assert!(out == cryptol_spec::merge_clocks(a, b));
    }

    #[crux_test]
    fn merge_vc_equiv() {
        merge_c_equiv_spec().enable();
        let a = <VectorClock>::symbolic("a");
        let b = <VectorClock>::symbolic("b");
        let out = merge_vc(a, b);
        crucible_assert!(out == cryptol_spec::merge_vc(a, b));
    }
}
\end{lstlisting}
\caption{
\label{fig:example-equiv}
Crux symbolic test proving that \lstinline{merge_vc} is equivalent to its
Cryptol specification. This symbolic test uses two steps of reasoning: first it
verifies a specification for \lstinline{merge_clocks}, and then it
verifies \lstinline{merge_vc} assuming the previous specification. 
}
\end{figure}

\Cref{fig:example-equiv} shows a Crux module which verifies the
functional correctness of the vector clock. Mathematically, this test
verifies the following properties: 
\begin{gather*}
  \forall a, b \in \mathtt{Clock}. \;
            \mathtt{merge\_clocks}(a, b)
    \equiv \mathtt{cryptol\_spec::merge_clocks}(a, b)
\\
\forall a, b \in \mathtt{VectorClock}. \;
            \mathtt{merge\_vc}(a, b)
    \equiv \mathtt{cryptol\_spec::merge\_vc}(a, b)
\end{gather*}

The symbolic test \lstinline{merge_vc_equiv} proves the
equivalence of the Rust implementation of \lstinline{merge_vc} to
the Cryptol specification
\lstinline{cryptol_spec::merge_vc(a, b)}. If we examine lines
12-19, note once again that this proof is similar in structure to a unit
test. The difference this time is that the test calls to a reference
specification, rather than just asserting a property of implementation
variables.

Scalability is a key concern when running Crux, because each symbolic
test is converted into a series of queries to the SMT solver. In fact,
the symbolic test \lstinline{merge_vc_equiv} could be verified as
a whole, without breaking up the verification task. However, for most
moderate-size verification problems, the SMT solver will not be able to
process the resulting verification conditions. To solve this problem, we
use \emph{compositional reasoning}, demonstrated here by the
\lstinline{merge_c_equiv} symbolic test.

Crux first verifies that the Rust function
\lstinline{merge_clocks} is equivalent to the Cryptol
specification \lstinline{cryptol_spec::merge_clocks(a, b)}. But
in addition, in any subsequent proof, it will replace any calls to
\lstinline{merge_clocks} with the executable specification code.
This is sound, because the two functions have been verified equivalent,
and because the specification is typically much simpler than the
implementation, it enables scalability.

We mark the test with \lstinline{#[crux_spec_for(merge_clocks)]}
to enable compositional reasoning about the
\lstinline{merge_clocks} function in other tests. The
implementation of compositional reasoning is explained in more detail by
\Cref{sec:architecture-comp}.

\subsection{A More Realistic Example: Chacha20}
\label{sec:example2}

Next we examine an example based on the ChaCha20 stream
cypher~\cite{bernstein2008chacha}. We target the implementation in
the Rust \lstinline{chacha20} crate~\cite{chacha20-crate}. For
illustrative purposes we only examine the signatures of the functions under test
and one functional correctness property: input-output correctness of the cypher
implementation function with respect to an executable specification.

\Cref{fig:chacha20-sigs-spec} top shows the Rust signatures from the
\texttt{chacha20} crate we will discuss. Lines 1-3 define the trait
\lstinline{Rounds} which is used to pass a statically known
constant via a phantom type to instances of the struct
\lstinline{Core} defined on lines 4-5. Our example focuses on the
methods of \lstinline{Core} defined on lines 7-9. To use this
interface a user first initializes a \lstinline{Core} by
statically passing a \lstinline{Round}-implementing type
parameter and then calling the \lstinline{new} method. Then to
compute part of the stream cypher the user calls
\lstinline{generate} which will call the private
\lstinline{rounds} function.

\begin{figure}
    \begin{lstlisting}
pub trait Rounds: Copy {
  const COUNT: usize;
}
pub struct Core<R: Rounds> { state: [u32; STATE_WORDS],
                             rounds: PhantomData<R>     }
impl<R: Rounds> Core<R> {
  pub fn new(key: &[u8; KEY_SIZE], iv: [u8; IV_SIZE]) -> Self {/*...*/}
  pub fn generate(&mut self, counter: u64, output: &mut [u8]) {/*...*/}
  fn rounds(&mut self, state: &mut [u32; STATE_WORDS]) {/*...*/}
}
    \end{lstlisting}

    \begin{lstlisting}
cryptol! {
  path "Primitive::Symmetric::Cipher::Stream::chacha20";
  pub fn cryptol_kexp(k: [u8; 32], c: u32, n: [u8; 12])
          -> [u32; 16] = "kexp";
  pub fn cryptol_core(x: [u32; 16]) -> [u8; 64] = "core";
  pub fn cryptol_cdround(x: [u32; 16]) -> [u32; 16] = "cdround";
}
    \end{lstlisting}

    \caption{
        \label{fig:chacha20-sigs-spec}
        \emph{Top:} Rust signatures found in the \texttt{chacha20} crate. \emph{Bottom:}
        Rust bindings to Cryptol in the \texttt{cryptol-specs} repository.
    }
\end{figure}

\Cref{fig:chacha20-sigs-spec} bottom shows the Rust bindings to a Cryptol executable
specification for ChaCha20 defined in a separate repository. Bindings in
the \lstinline{cryptol!} macro may reference names in the
imported module or contain inline Cryptol code. Here we only bind the
external names to Rust names and type signatures.

Our compositional verification of ChaCha20 begins in
\Cref{fig:chacha20-test1} where we show that the implementation's
\lstinline{rounds} function on instances of
\lstinline{Core} is extensionally equal to ten calls of the
executable specification's \lstinline{cryptol_cdround} function.
Line 1 declares that this test function is a Crux-MIR \emph{spec} for
calls to \lstinline{rounds} on instances of
\lstinline{Core} which do twenty rounds (the definition of that
instance of the phantom type is not shown). Since this test function is
declared a Crux-MIR spec, we can use it to reduce the computational
burden of verifying the caller of \lstinline{rounds}. Line 5 is
an \emph{assumption} about the symbolic state variables declared on
lines 3-4, namely, that the two are equal. Lines 7-9 run the
implementation on the symbolic state values by instantiating a
\lstinline{Core} struct value and calling the
\lstinline{rounds} function. Next, we run the executable
specification on lines 11-17. Since the
\lstinline{cryptol_cdround} function does a little less work than
our implementation, the test performs those extra steps. Finally, lines
19-21 \emph{assert} that the two resulting states are equal, just as you
would in a unit test.

\begin{figure}
    \begin{lstlisting}
#[crux_spec_for(Core::<R20>::rounds)]
fn chacha20_rounds_equiv() {
    let state1 = <[u32; 16]>::symbolic("state1");
    let state2 = <[u32; 16]>::symbolic("state2");
    crucible_assume!(state1 == state2);

    let mut rust_core = Core { state: state2, rounds: PhantomData };
    let mut rust_state = state2;
    Core::<R20>::rounds(&mut rust_core, &mut rust_state);

    let mut cryptol_state = state1;
    for _ in 0..10 {
        cryptol_state = cryptol_cdround(cryptol_state);
    }
    for (a, b) in state1.iter().zip(cryptol_state.iter_mut()) {
        *b = b.wrapping_add(*a);
    }

    for (&x, &y) in rust_state.iter().zip(cryptol_state.iter()) {
        crucible_assert!(x == y);
    }
}
    \end{lstlisting}
    \caption{
        \label{fig:chacha20-test1}
        Crux-MIR spec for an internal method in the ChaCha20
        implementation.
    }
\end{figure}

We complete verification of this part of ChaCha20 in
\Cref{fig:chacha20-test2} by showing that the implementation's
initialization call to \lstinline{new} followed by a call to
\lstinline{generate} is extensionally equal to calling the
executable specification's \lstinline{cryptol_kexp} followed by
\lstinline{cryptol_core}. Line 3 enables compositional reasoning
about uses of the function \lstinline{rounds}, line 5 declares
three symbolic inputs by leveraging Rust's normal type inference. We run
the implementation on lines 7-9 by passing in \lstinline{key} and
\lstinline{iv} to the initializer, \lstinline{new}, and
calling \lstinline{generate} on \lstinline{counter}.
Similarly, we run the executable specification on lines 11-12, and last
assert that their outputs are equal on lines 14-15.

\begin{figure}
    \begin{lstlisting}
#[crux_test]
fn chacha20_correct() {
    chacha20_rounds_equiv_spec().enable();

    let (key, iv, counter) = <_>::symbolic("inputs");

    let mut rust_core = Core::<R20>::new(&key, iv);
    let mut rust_output = [0; 64];
    rust_core.generate(counter, &mut rust_output);

    let cryptol_round = cryptol_kexp_adjusted(key, counter, iv);
    let cryptol_output = cryptol_core(cryptol_round);

    for (&x, &y) in rust_output.iter().zip(cryptol_output.iter()) {
        crucible_assert!(x == y);
    }
}
    \end{lstlisting}
    \caption{
        \label{fig:chacha20-test2}
        Crux-MIR symbolic test using the spec in \Cref{fig:chacha20-test1} to
        verify the interface for ChaCha20.
    }
\end{figure}

\section{Crucible Symbolic Execution Library}
\label{sec:technical-crucible}




Crucible~\cite{crucible} is a strongly-typed imperative language intended as a 
translation target for other languages. Crucible lies at the heart of SAW-Cryptol 
and Crux. It is built on an underlying library called What4~\cite{what4} that provides 
connections to a variety of SAT and SMT solvers. 

Crucible supports many data types found in SMT solvers including
primitive types (booleans, bit-vectors, integers, arrays), aggregate
types (tuples, structs, vectors, enums/tagged unions), as well as
symbolic values of those primitives and aggregates. Variables may be
local or global, and Crucible includes notions of mutation. Since it is
intended to model other languages, Crucible has no syntax of its own,
however it is designed to be extended.

Users may implement \emph{Crucible language extensions} in Haskell to
complete a Crucible model of a desired language. For example, extensions
may be used to add a memory model to Crucible, which does not include a
memory model of its own. New primitive types, such as inductive
datatypes or pointers, and new primitive operations, such as pointer
dereferences, may also be added via extension. Crucible's own primitive
types are implemented as a Haskell datatype designed to make adding such
primitive types simpler.

The \emph{Crucible symbolic execution engine} runs a program via symbolic
execution. Crucible's strategy involves cloning the symbolic state at branch 
points and then merging back together at
post-dominator nodes in the control flow graph, in an attempt to build a single
symbolic state representing the result of a program. As such, it can be thought 
of as a translator from imperative programs to mathematical
formulas.

The result is an environment which maps program variables to symbolic
values (expressions over symbolic variables) and a set of assumptions and
assertions which the symbolic variables must satisfy. These expressions reflect
everything that happens to a program variable, such as assignments, branching on
conditions over other program variables, and primitive operations requiring
bit-precision. Loops are symbolically executed by repeatedly applying the loop 
body until the SMT
solver reveals that the loop condition cannot hold. 
In Crux-MIR, all \lstinline{panic!}
invocations in the program and its libraries are treated as failed assertions,
so Crux-MIR also attempts to prove that the program does not panic.

Crucible represents formulas internally as directed acyclic graphs (DAGs) to
help keep them at a manageable size when the same subformula appears repeatedly.
It also performs automated term simplification of terms as they are built, and
performs simple abstract interpretation in parallel with symbolic execution to
track, for instance, the possible ranges of program variables.

\section{Crux-MIR Architecture}
\label{sec:architecture}

Crux-MIR first extracts an intermediate representation of the
Rust code using a hook in the normal compilation process
(\Cref{sec:architecture-mir}). Then it compiles the Rust intermediate
representation to a Crucible program (\Cref{sec:architecture-crucible}).
Last, it issues an SMT query to prove that the property holds
(\Cref{sec:architecture-smt}). Additionally, compositional reasoning
requires specific behaviors at each stage
(\Cref{sec:architecture-comp}).

\subsection{Capture of mid-level IR}
\label{sec:architecture-mir}

The first step is to extract a representation of the code in the form of MIR,
the Rust compiler's Mid-level IR~\cite{mir}. The MIR representation at a higher 
level of abstraction than the corresponding LLVM IR
produced by a later stage of the Rust compiler. Crux-MIR analyses the Rust
program at this level because it allows a higher level of abstraction in the
corresponding Crucible program, excludes some cases possible to express in LLVM
IR but not in MIR, and leads to more efficient use of SMT.

Extraction is performed by a custom tool,
MIR-JSON, which links against the Rust compiler as a plugin. MIR-JSON
pre-processes the input Rust code and emits a JSON-formatted representation of
MIR code to a file before the Rust compiler continues with its normal
compilation pipeline. By default, MIR-JSON also runs a script to extract the MIR
representation of the Rust standard library, which is used by the Crux-MIR to
generate Crucible as needed by users' code.

MIR-JSON is designed to preserve Rust semantics of users' code while
producing a representation which admits tractable analysis. Several
trade-offs are made in favor of efficient analysis in Crux-MIR. 

Lifetimes are erased by MIR-JSON before extraction so that Crux-MIR
does not need to reason about reference lifetimes or whether
references are used safely. It is assumed that the Rust borrow
checker's analysis is correct.

There is differing support for generics in Rust and Crucible. For
example, in Rust it is possible to reference the associated types of a
type variable constrained to a specific trait, but Crucible does not
allow this use of a type variable. MIR-JSON calls internal functions of the Rust
compiler to monomorphize both data and function definitions and
produce only concrete, non-generic MIR code for Crux-MIR to ingest.\footnote{MIR-JSON
walks the AST at each crux-test entry point, looking for generic
functions called on inferred type parameters and emits a
monomorphized version of that function at the type, until all calls
are monomorphized.} To implement dynamic dispatch, trait objects are
compiled to use tables of function pointers.

\subsection{Compilation to Crucible}
\label{sec:architecture-crucible}

Crux-MIR next compiles MIR into a Crucible program. The types found in MIR code
mostly translate straightforwardly into Crucible's primitive types, and the
control flow graph in MIR is preserved unchanged in the Crucible output. The
result is an abstract representation of the original Rust code's semantics, with
a well-defined SMT-LIB semantics against which we can discharge queries.

\paragraph{Memory model.}

Crux-MIR represents each Rust memory allocation as a separate mutable location and
reimplements primitive read and write operations against this
representation. Allocations are strongly typed, and most Rust
types translate easily into Crucible's primitive types. Rust structs are
represented as Crucible structs and Rust vectors are represented as a
Crucible struct with a backing array.

Crux-MIR extends Crucible's memory model to represent Rust references
into this memory. Rust references are implemented as a
pair of an allocation and a path to a
sub-component. A path may include both struct fields and array offsets to
locate the component held in the Rust reference.

The simplified memory model used by Crux-MIR presents some limitations.
Crux-MIR does not expose the concrete representations of types in terms of flat
arrays of bytes, so it generally does not support operations that rely on a
particular layout or reinterpret the bytes of a value as a different type.
Such unsupported operations include the \lstinline{mem::transmute} function,
use of C-style unions, casts between different pointer types, and pointer
arithmetic outside of arrays.
These limitations prevent Crux-MIR from directly handling some types of unsafe
Rust code.

\paragraph{Implementation of intrinsics.}

Crux-MIR implements Rust intrinsics by
encoding their semantics directly in Crucible.
Most Rust intrinsics are not directly exposed in the standard library:
instead wrappers dealing with concrete types are exposed. Crux-MIR
directly implements some intrinsics, and for others it overrides the wrappers
exposed by the Rust standard library. These definitions provide
good coverage of the Rust standard library for most programs.

Some definitions for intrinsics used by Crux-MIR are less versatile than their
Rust counterparts. For example, we implemented
transmute\footnote{\lstinline{std::mem::transmute<T, U>(e: T) -> U}}
to perform coercions between between primitive integer
types and byte arrays only. Since this \lstinline{transmute}
doesn't perform arbitrary conversions between types, many instances of
\lstinline{unsafe} code used in Rust are not supported by Crux-MIR
and will fail during compilation. However, since the use of
\lstinline{unsafe} is limited in practice, this 
does not present a problem for our use-cases.

\paragraph{Standard library.}

Crux-MIR overrides some parts of the Rust standard library to work with our
memory model or admit more efficient analysis by SMT. These definitions are
either written directly using Crucible constructors in Haskell or are provided
as alternative Rust implementations. However, with our current memory model
significant portions of the Rust standard library (eg.
\lstinline{std::vec}) require no redefinition to work with Crux-MIR
because they depend on lower-level components which we have overridden. The
distance between Crux-MIR's memory model and the real memory model of Rust
largely determines the amount of definitions that we must provide. 

One example is Crux-MIR's simplification of the Rust standard library
function for in-place stable sort of slice values,
\lstinline{slice::sort(&mut self)}. This function uses \emph{``an
adaptive, iterative merge sort inspired by timsort''} which analyses the
slice contents and chooses a different strategy depending on input 
characteristics.
For most purposes, the only invariants a user cares about is that the result 
is stably sorted.
Therefore, to simplify the SMT expression generated by code which uses
\lstinline{slice::sort}, and achieve shorter verification times,
Crux-MIR uses a simple insertion sort algorithm. 

\subsection{SMT verification of properties}
\label{sec:architecture-smt}

Crux-MIR generates an abstract representation of a user's Rust code in SMT-LIB
in order to check whether user-supplied assertions about the code hold. 
Intuitively, Crux-MIR uses Crucible to produce equations that
describe what is known about variables after a program is run, and SMT
is used to find an assignment of values to those variables which obeys
those equations but falsifies the user assertions. If no such
counterexample to the user assertions is found, then the properties they
represent hold. If an assignment is is found, Crux-MIR 
outputs an assignment of variables which would violate the desired
property.

The full sequence is: (1) Crux-MIR compiles the Rust code to Crucible
(\Cref{sec:architecture-crucible}). (2) The Crucible symbolic execution engine
runs to obtain SMT-LIB assertions (\Cref{sec:technical-crucible}). These are
equalities on program variables and SMT expressions written in terms of past
state. (3) Crux-MIR adds assertions to represent assumptions about the initial
state of program variables. (4) Crux-MIR adds assertions which are the negation
of user-supplied assertions on program variables, to represent the property we
seek to verify. (5) Finally, the SMT solver executes and searches for an
assignment which satisfies all of the assertions.

\subsection{Compositional reasoning}
\label{sec:architecture-comp}


Compositional reasoning is necessary to verify even mid-size pieces of code. As
shown above, compositional reasoning in Crux-MIR works by splitting a large
symbolic test into a \emph{spec} and a symbolic test that uses the spec. Further
splitting of specs into yet more parts, each of which can be checked quickly,
scales verification to larger projects. Each spec need only reference its direct
descendants. 

To gain an intuition of how compositional reasoning is implemented,
let us discuss an extensional equality test. If a user wishes to show the
equality of implementation \(I(v)\{ \dots \}\) to reference
implementation \(R(v)\{ \dots \}\), but the test takes too long to
check, they may apply the following technique:

\begin{enumerate}
\def\labelenumi{\arabic{enumi}.}

\item
  Factor out subexpressions of \(I(v)\{ \dots i(v) \}\) and
  subexpressions of \(R(v)\{ \dots r(v) \}\) and show the equality of
  subexpressions \(i\) to \(r\). This test showing the equality of \(i\)
  to \(r\) is a \emph{spec} of \(i\) and so the user marks it as a spec.

\item
  Next, in the test showing equality of \(I\) to \(R\), enable the spec
  of \(i\).

\item
  When checking equality of \(I\) to \(R\), Crux-MIR finds that \(i\)
  has a spec enabled and replaces calls to \(i\) with \(r\). The task of
  proving equality of \(I\) to \(R\) is reduced to showing equality of
  \(I(v)\{ \dots r(v) \}\) to \(R(v)\{ \dots r(v) \}\) which only
  compares the parts elided by the ellipses.
\end{enumerate}

Crux-MIR constrains a test marked as a \emph{spec} to contain only one
call to the function under test. After verifying assertions in the test
in the normal way (described in \Cref{sec:architecture-smt}), Crux-MIR
performs additional steps to support compositional reasoning. Crux-MIR
extracts the \emph{most general implementation} that would satisfy the
specification expressed by the test.
This extracted implementation will be used to replace calls to the function under test in any
tests which enable this spec. For example, Line 3 of \Cref{fig:example-comm},
\lstinline{#[crux_spec_for(merge_clocks)]}, marks the test
\lstinline{fn merge_c_equiv} as a spec for the function
\lstinline{fn merge_clocks}.

A user may \emph{enable a spec} in a test to gain the efficiency benefit
promised by compositional reasoning. When a spec is enabled in a test,
Crux-MIR replaces calls to the indicated function with an arbitrary
symbolic value constrained to have the property proved by that spec. If
the spec is an extensional equality property then Crux-MIR replaces calls to the
indicated function with calls to the equivalent implementation. 

For example, Line 14 of \Cref{fig:example-comm},
\lstinline{merge_c_equiv_spec().enable();}, enables the spec of function
\lstinline{fn merge_clocks} in the test
\lstinline{merge_vc_equiv}. Now when checking the assertions in a test
which uses a spec, the SMT-LIB assertions required by the spec are trivially
satisfied for the relevant subexpressions. This reduces the verification task to
analysis of the code not covered by the spec.


\subsection{Cross-language support}

Crux supports two specification languages, hacspec and Cryptol. hacspec is an
executable specification language defined as purely functional subset of
Rust~\cite{hacspec}~\cite{hacspec}. 
Crux-MIR can work with specifications written in hacspec
because the are valid Rust, so our Rust support can serve double duty as
a specification idiom as well.

Cryptol is an executable algorithm-specification language designed for
specifying, testing, and verifying cryptography
algorithms~\cite{cryptol,cryptolpaper}. Crux-MIR is uses the SAW-core libraries
to interface with specifications written in Cryptol by converting to SMT-LIB
expressions.

Crux-MIR supports verifying Rust, Cryptol, and hacspec code against each other
by converting each language to SAW-Core, which is the common interchange
language of the SAW-Cryptol toolchain. Much of the required functionality,
including symbolic execution support for Cryptol, was already present in SAW;
when developing Crux-MIR, we simply called upon existing functionality,
adding a relatively small amount of Rust-specific conversion code. This is
one way in which building on an existing toolchain makes creating new tools
easier. 

\subsection{Tracking MIR versions}

MIR changes as Rust evolves, and as a practical matter, these changes
require upkeep to our tools. Many changes are
introduce cosmetic adjustment, such as using two fields where a tuple
was previously used, renaming fields or constructors, or adjusting where
casts are applied. Some changes are more fundamental, and
require significant effort to update either MIR-JSON or Crux-MIR. 


For example, over time Rust has changed to perform more constant evaluation at
compile time. A Rust interpreter was added to the Rust compiler, which
allows constant expressions and referencing code to be partially
evaluated. Some \lstinline{const} declarations
are no longer represented in MIR as a primitive value, but instead as a
byte array. To analyze values represented this way, Crux-MIR or the
Crucible program it produces must interpret the byte array
appropriately. For example, it is sometimes necessary to reify a data
structure represented by raw bytes at compile time (such as manually
constructing a \lstinline{(Int, Ptr)} tuple) to output a high
level Crucible data structure. Recent MIR versions push the abstraction
level of the MIR language even lower, allowing very complex expressions
in const expressions, but requiring more upkeep in Crux-MIR.

\section{Case studies}
\label{sec:case-studies}

\subsection{Example 1: Shardstore serialize/deserialize}
\label{sec:shardstore}

As briefly described in Bornholt \emph{et al}~\cite{shardstore}, Crux-MIR has previously been
applied by a team at Amazon Web Services as part of the assurance process for
Shardstore, a key-value storage node implementation. Crux was used to 
\begin{quote}
{\it  
[...] prove panic-freedom of our deserialization code. We proved that for any sequence
of on-disk bytes (up to a size bound), our deserializers cannot panic (crash the
thread) due to out of bounds accesses or other logic errors.} \cite[p.10]{shardstore}.
\end{quote}

We note that deserialization code is not inherently
bounded, unlike many cryptographic primitives. This is why a size bound is
necessary.


\begin{figure}[t]
    \centering
    \includegraphics[width=\textwidth]{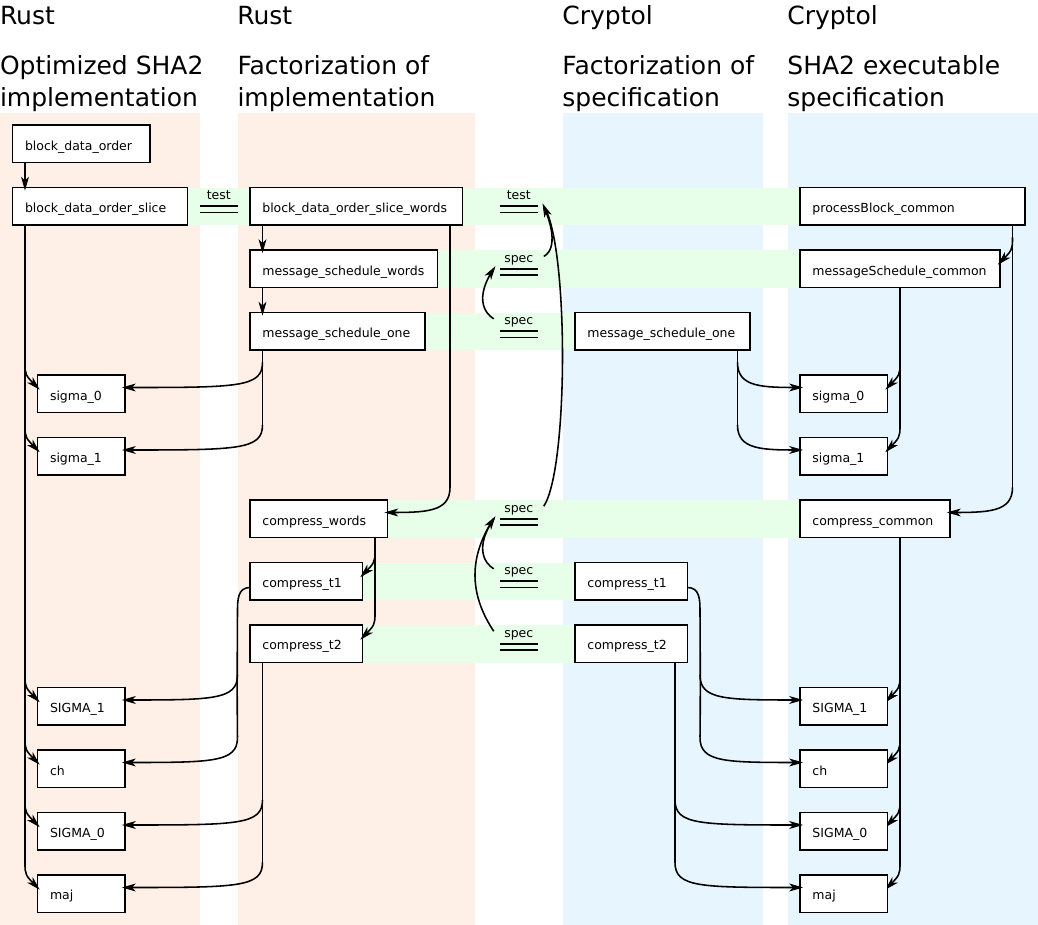}
    \caption{
        \label{fig:sha2-rust-vs-cryptol}
        Structure of our Crux-MIR verification of a Rust implementation of SHA2
        against an executable specification written in Cryptol.
        Functions are represented by white boxes and function calls by downward
        arrows.
        Crux-MIR symbolic tests and specs are shown as large equalities between function
        boxes.
        The use of compositional reasoning is represented by upward arrows
        leaving Crux-MIR spec equalities.
    }
\end{figure}


\subsection{Example 2: \texttt{ring} SHA1 and SHA2}

We have verified Rust implementations of SHA1 and SHA2 from the \texttt{ring}
crate~\cite{ringlibrary} against generic (i.e. not Rust or
\texttt{ring}-specific) specifications taken from the
Cryptol~\cite{cryptol-spec-repo} and hacspec~\cite{hacspec-spec-repo}
archives of standardized specifications.

Each proof required several layers of symbolic tests combined using
compositional reasoning. In each of the proofs, a top level Crux-MIR symbolic
test shows that the block-processing function generates the same outputs the
specification, when given the same inputs. This proof structure is standard for
Crux and mirrors that of 
\lstinline{crux_spec_equivalence} in the vector clock example
(\Cref{fig:example-equiv}). 

In complex proofs it is common to prove extensional equality over multiple
steps. This exploits the transitive nature of equivalence. For example, in the
SHA2 proof of equivalence to Cryptol we defined an alternate \emph{factorized}
implementation of the top level function,
\lstinline{block_data_order_slice}. This contains nearly the same code
but it is broken into separate functions that isolate components, e.g.~a loop or
a loop body. We also factorized some parts of the Cryptol specification to line
up with the factorized implementation. The proof then establishes equivalence in
several steps: from optimized Rust, to factorized Rust, to specification.
\Cref{fig:sha2-rust-vs-cryptol} shows this proof structure in diagrammatic form.

Factorization often requires careful design work, rather than computational
cost. For example single Crux-MIR symbolic test was sufficient to prove the
equivalence of \lstinline{block_data_order_slice} with the top level
function of our factorized implementation,
\lstinline{block_data_order_slice_words}, because the two implementations
differ only in the presence of additional function boundaries. Much of the proof
engineering effort was spent identifying which subcomponents could be factored
out in common. We frequently consulted the FIPS-180 standard~\cite{SHA} to
understand the Rust implementation and the Cryptol specification. 


As an example of factorization, consider the function
\lstinline{compress_words} in \Cref{fig:sha2-rust-vs-cryptol}.
Crux-MIR cannot verify \lstinline{compress_words}
equivalent to the Cryptol \texttt{compress\_common} function directly, because both
perform a loop which results in a too-large SMT-LIB expression. We defined
\texttt{compress\_t1} and \texttt{compress\_t2} to capture the effect of a
single iteration of the loop. These functions are each only moderately
complex so they can be verified easily. When these specs are enabled, the loop bodies 
for \lstinline{compress_words} and
\texttt{compress\_common}
are trivially equal, and can be verified
automatically. 

Logistically, an engineer with a short introduction to Crux-MIR and no
background in cryptography was able to complete the verification of SHA1
against its Cryptol specification in three weeks. Following that,
verification of SHA1 against hacspec, SHA2 against Cryptol, and SHA2
against hacspec took an additional three weeks. Each of the four proof
developments take under 10 minutes to re-verify on a recent MacBook Pro.

\section{Related work}
\label{sec:relatedwork}

The immediate predecessor to Crux is SAW-Cryptol~\cite{saw,cryptol} and the proofs
accomplished using these
tools~\cite{continuous-verification-s2n,Boston2021,DBLP:journals/ieeesp/Dodds22}.
Crux builds on the same underlying toolchain and in the long run, Crux is
intended to fill a similar role to SAW-Cryptol with an interface that is easier
to learn and use. Crux supports most of the features that would be necessary to
replicate the AWS-LibCrypto proofs, the most challenging proofs undertaken using
SAW-Cryptol so far~\cite{Boston2021}. The main missing feature is support for
SAW-Core term rewriting, which is necessary to discharge the most complex goals.
The two tools also cover distinct but overlapping sets of languages---Crux
supports Rust and LLVM, while SAW-Cryptol supports LLVM, x86 assembly, Java, and
several other languages. 

Several libraries in the SAW-Cryptol toolchain have been informed by previous
projects. For example, Crucible has similarities to the Boogie
verification language~\cite{DBLP:conf/fmco/BarnettCDJL05}, and the solver
interface library What4 was influenced by Why3~\cite{boogie11why3}. Crucible,
What4, and related libraries have been highly tuned for
performance and predictability in our domain of intricate bounded 
cryptographic code. 

Crux, like SAW-Cryptol before it, targets code that has not been
designed for verification. In parallel, projects such as HACL* have
developed from-scratch verified cryptographic
libraries~\cite{hacl-star}. We see these projects as complementary: HACL* and
similar are useful for new-build systems, while Crux targets the much larger
body of legacy code. See Dodds for a discussion of this
tradeoff~\cite{DBLP:journals/ieeesp/Dodds22}. 

There are now a significant number of Rust formal verification tools. We only
cite those that have close similarities to Crux-MIR---see Reid for a broader
informal survey conducted in  2021~\cite{rust-verification-survey-2021}. Several
Rust tools support compositional reasoning in some form, including
Prusti~\cite{prusti} and Aeneas~\cite{aeneas}. Aeneas appears closest---like
Crux, its stated objective is to allow mostly-automated verification---but
Aeneas presents a very different proof interface where Rust's memory safety
guarantees are leveraged to extract functional models of code. Crux does not
leverage Rust's region-based type system, which has two effects: we are able to
handle both Rust unsafe code, but we can only handle bounded memory structures.
Previously, the Heapster tool extended SAW-Cryptol with a similar type-based 
extraction mechanism to Aeneas~\cite{heapster}. 

The most similar tool to Crux-MIR in terms of proof interface is
Kani~\cite{kani}. Both Crux-MIR and Kani are bit-precise verification tools
where proofs are written as symbolic tests---Kani inherits this specification
approach from its predecessor, CBMC~\cite{cbmc}. The main differences appear to
be driven by different use-cases. Crux targets proofs of highly intricate 
code, which requires that we support compositional reasoning, and
proofs of extensional equality to executable specifications. Kani is targeted at
simpler inline properties written as Rust assertions, and as such does not
support compositional reasoning or external specification languages.

\section{Conclusion}

We have described Crux, a cross-language verification tool, and its 
Rust instantiation Crux-MIR. Crux provides a new interface to a
technology with proven effectiveness in industry: compositional symbolic
simulation in the SAW-Cryptol toolchain. We and others have demonstrated Crux on pieces
of real-world Rust code. Crux also demonstrates 
how tools can be built using the SAW-Cryptol ecosystem, in particular 
thanks to the Crucible library for symbolic simulation. 

\subsubsection*{Acknowledgements}

Crux has been built by a large team 
over a long period of time, with several applications in mind. 
As such, many people have influenced its design. 
We particularly thank 
  James Bornholt, Rajeev Joshi, Serdar Tasiran and Stephanie Weirich. 
We also thank the many other authors of the SAW-Cryptol toolchain over
twenty years. 

%
%

\clearpage 

\bibliographystyle{splncs04}
\bibliography{mybibliography}

\end{document}